\documentstyle[11pt,epsf,twoside,jltp]{article}

\title{Quantum Relaxation of Ensembles of Nanomagnets}

\author{N. V. Prokof'ev$^{1}$  and P. C. E. Stamp$^{2,3}$}
\address{
$^{1}$ Russian Research Center "Kurchatov Institute", Moscow 123182, Russia\\
$\;\;\;$ \\
$^{2}$ Department of Physics and Astronomy, and Canadian Institute for
Advanced Research,
University of British Columbia, 6224 Agricultural
Rd., Vancouver B.C., Canada V6T 1Z1 \\
$^{3}$ Institut Laue-Langevin, Ave. des Martyrs, Grenoble 38042, France \\
}
 
\runninghead{N.V. Prokof'ev and P.C.E. Stamp}{Quantum Relaxation of 
Ensembles of Nanomagnets}

\begin{document}

\begin{abstract}
Recent theory and experiment in crystals of molecular magnets suggest that
fundamental tests of the decoherence mechanisms of macroscopic quantum
phenomena may be feasible in these systems (which are almost ideal
quantum spin glasses). We review the results, and suggest new experiments.
\end{abstract}

\maketitle

\vspace{3mm}

The attempt to push quantum effects to mesoscopic
or macroscopic scales has led, in magnetic systems, to
various interesting discoveries. Theoretical work on
domain wall tunneling \cite{Stamp91,Tatara,Dube} led to experiments
claiming to see tunneling of single domain walls \cite{4}. Early
theory on tunneling of large spins \cite{Hemmen,Enz}
stimulated experiments on resonant tunneling in crystals of magnetic
macromolecules
\cite{QTM,Novak,Barbara,Friedman,Thomas,Sangregorio},
and on ``macroscopic quantum coherence" \cite{Awsch} in
giant ferritin molecules.

There is a theoretical complication - the environmental ``spin bath" of
nuclear and paramagnetic spins should destructively interfere with any
tunneling \cite{Stamp88,PS93,PS95,Garg}. In fact, except in ferritin,
all experiments see {\it incoherent} tunneling relaxation at low T.
A detailed theory of such relaxation,
\cite{PS96,TPS97,T98,PS98}, applied to ensembles of tunneling
nanomagnets, makes several ``universal" predictions for the relaxation
characteristics, some of which have now been verified
\cite{Ohm98,Thomas98}. Here we discuss the recent developments, and
suggest future experiments designed to bring out the fundamental role of
nuclear spins in quantum decoherence at low $T$.

\section{Kinetics of Quantum Relaxation}

Our problem is simple but general. Consider an array (ordered or otherwise)
of identical
2-level systems (which here represent large spins; $S \gg 1$, below the
crossover
temperature $T_c$ of confinement to their 2 lowest levels). Their
Hamiltonian is \cite{PS98}
\begin{equation}
H = {1 \over 2} \sum_{i\ne j} V^{d}_{ij} \hat{\tau}_z^{(i)}
\hat{\tau}_z^{(j)}
+ \sum_j \Delta  \hat{\tau}_x^{(j)}
+\sum_{jk} V_N (\hat{\tau}_z^{(j)} , \{ \vec{I}_k \} )
+ H_{NN} \;,
\label{Hamiltonian}
\end{equation}
where $V^{d}_{ij}$ describes dipolar interactions between nanomagnets
$\{ i,j\}$, $\Delta $ is the tunneling amplitude between states
$\alpha =\pm $, $V_N$ couples nanomagnets to nuclear spins $\{
\vec{I}_k \} $, and $H_{NN}$ describes the {\it intrinsic}
nuclear spin dynamics.
Note that without $V_N$, Eq.~(\ref{Hamiltonian}) describes
the ``quantum spin glass" problem \cite{23}; the effect of $V_N$ is to
couple this to an environment.

The entire system is
described by normalised probability distributions for single
molecules, pairs of molecules, etc.; thus $P_{\alpha } (\xi ,\vec{r}
; t ) $ is the probability that a molecule at position $\vec{r}$
has polarisation $\alpha = \pm$ and bias $\xi = \epsilon_+ - \epsilon_-$
between states $\vert \pm \rangle \equiv \vert \pm S \rangle$.
The bias $\xi$ at $\vec{r} = \vec{r_i}$ sums the nuclear bias (of typical
energy scale $E_0$), the dipolar bias
$\sum_j V^d_{ij} \tau_z^{(j)}$ (typical energy scale $E_D$),
and any bias from external fields. This 1-particle distribution
$P_{(1)}$ is connected to higher distributions $P_{(2)}, P_{(3)}$,
etc., via a BBGKY- style hierarchy of kinetic equations \cite{PS98}.
Since the only part of the environment with any dynamics at
low T and low H (ie., able to mediate irreversible incoherent tunneling)
is the nuclear bath, the
kinetic equation for $P_{(1)}$
is governed by a nuclear-mediated transition rate $\tau_N^{-1}$, calculated
in \cite{PS96}. When the number of nuclear spins
co-flipping with the molecular spin is small,
\begin{equation}
\tau_N^{-1}(\xi ) \sim \tau_0^{-1} e^{-\vert \xi \vert / \xi_o }\;,
\label{tauN}
\end{equation}
where $\xi_o$ parametrises the range of fluctuations in $\xi$, and
$\tau_0^{-1} \sim \Delta^2/\xi_o$; typically $\Delta \ll \xi_o \ll E_0, E_D$.

Several unambiguous predictions \cite{PS98} flow from the kinetic
equation.
Over a wide time range ($E_D/\xi_0 > t/\tau_0 > \xi_0 /E_D$) the 
magnetisation should relax from saturation
according to a {\it universal square-root law},
i.e., $M(t) =M_0[1-(t/\tau_{Q})^{1/2} ]$. For ellipsoidal samples
$\tau_Q \rightarrow \tau_Q^{ell} = (E_D\tau_0 /\xi_0 ) f(c)$ (where
$f(c)$ is an analytic function of the ellipsoid shape);
for other shapes
$\tau_Q \sim \tau_0^{ell}/\xi_0N_{\uparrow}(0)$,
where $N_{\uparrow}(0)=\int d^3r P_{\uparrow}(\xi =0,\vec{r}, t=0)$ is the
initial ``up" distribution at zero bias.
Since $\tau_Q^{ell} \sim E_D/\Delta^2$, we have
$\tau_Q \sim E_D^2/\Delta^2\xi_0$ for a generic shape
(since $N_{\uparrow}(0) \sim O(1/E_D)$). In Figure 1 we see why the square root
prediction is valid independently of the details of the
hyperfine couplings, the sample shape, or the ratio $E_0/E_D$; it arises
from the "Lorentzian hole", eaten in the initial $N_{\uparrow}(\xi)$ by the
decay, coming from the long-range dipolar interactions
\cite{PS98}. Fig. 1 shows a Monte Carlo simulation 
for a cube of size $(75)^3$ sites, and depicts 
$N_{\uparrow}(\xi)$ when $t=0$ (so $M=M_0$), when $M/M_0$ has relaxed
to 0.99; and the difference between the two.
\begin{figure}[t]
\epsfxsize=5.5in
\epsfbox[15 250 580 580]{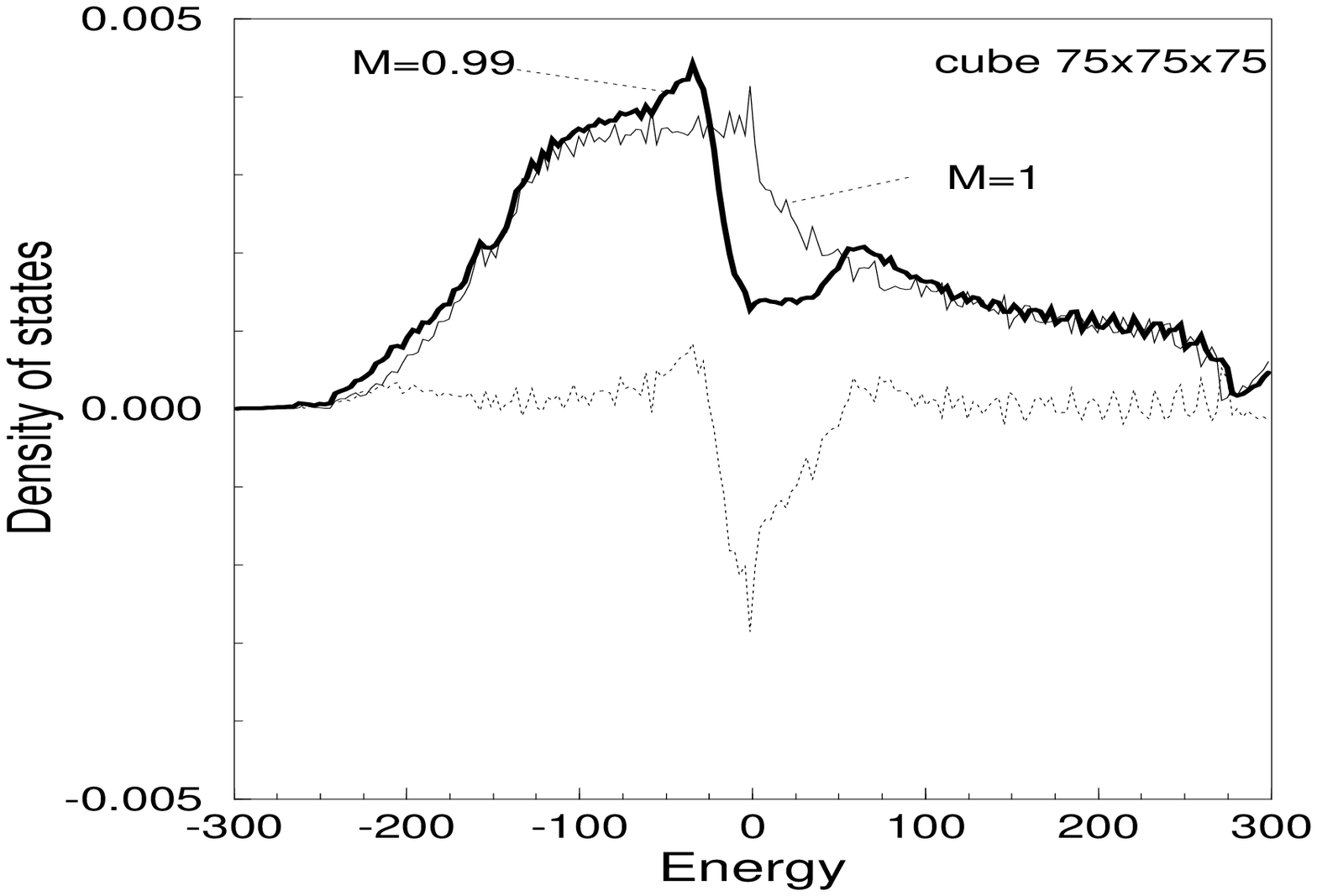}
\caption{}
\label{pphase}
\end{figure}
In a weak external field, we must replace
$N_{\uparrow}(0)$ in the expression for $\tau_Q^{-1}$
by $N_{\uparrow}(\xi = g\mu_B SH)$;
thus a field scan will find that
$\tau_Q^{-1} \propto N_{\uparrow}(\xi = g\mu_B SH)$, ie., 
$\tau_Q^{-1} (H)$ directly measures $N_{\uparrow}(\xi)$. Since
$N_{\uparrow}(\xi )$ can be calculated numerically (knowing $E_0,E_D$, and the
sample shape), this is a testable prediction.

\section{Universality and Crossovers}

How "universal" is the square root law? One potential problem, given that the
ligand groups in such large molecules can be unstable, is rogue fast-relaxing
"impurity" molecules.
Suppose the impurity concentration is $\delta_0 \ll 1$. Defining
$M_A(t) =(1-\delta_0) -x(t)$, and $M_B(t)= \delta_0-y(t)$,
for the bulk and
impurity magnetisations, normalising  $M_0=1$, and assuming a ratio
$\Lambda \gg 1$ between impurity and bulk relaxation rates, one gets a short-time
behaviour
\begin{equation}
y(t) =  \delta_0 \left[ 1-
\exp \left( -\Lambda x(t)/(1+\delta_0) \right) \right]\;,
\label{imp-1}
\end{equation}
\begin{equation}
t/\tau_Q^{A} =x^2(t) +2\delta_0 \left[
x(t)+{1+\delta_0 \over \Lambda } \left(
\exp \left( -\Lambda {x(t) \over 1+\delta_0 } \right) -1 \right) \right]
\:.
\label{imp-2}
\end{equation}
One easily verifies that for arbitrary $\Lambda \delta_0$,
any appreciable impurity contribution to $dM/dt$ destroys
square-root relaxation (conversely, square-root relaxation only appears
if the impurity contribution is negligible).

The square root behaviour must also break down in the crossovers to
long-time relaxation, and to thermal
activation.
After  $\sim 10-15 \%$ of the sample has relaxed, the Lorentzian hole
will distort because of intermolecular correlations.
These are not because of multi-flip processes (which have rate
$\sim  \Delta^4 /\xi_0^2E_D$, and so are rare), but because factorisability
of $P^{(2)}$ fails. However at long times other universal
results emerge. {\it First}, if hyperfine spread
in $N_{\uparrow}(\xi )$ is strong (ie., when
$E_0 > E_D$), then at times $>T_1(T)$, the $T_1$ nuclear relaxation
will sweep $\xi (t)$ at any molecule over most of the range $\Delta \xi$
of $N(\xi )$;
one then gets \cite{PS96} $M(t) \sim e^{-t/T_1(T)}$, for $t \gg T_1$.
For weak hyperfine coupling ($E_0 \ll E_D$),
intermolecular correlations give a complicated quantum
spin-glass problem, with no "sweeping" mechanism to
wipe out these correlations.
{\it Second}, if we thermally recycle the sample, depolarising at high $T$
until $M(t)/M_0 \ll 1$, the remaining relaxation will be exponential,
at a rate \cite{PS98} (with $\kappa \sim 1$):
\begin{equation}
\tau_{long}^{-1} \sim {2 \xi_0 \over
\tau_0 E_D [1+\kappa \ln (E_D/\pi\xi_0 )] } \;,
\label{tauL}
\end{equation}

The quantum-thermal crossover is more subtle. The {\it spin gap} in the system
is $\Delta \omega = (E_S - E_{-(S-1)})$, between the lowest levels
$\vert \pm S \rangle$ and
the next 2 levels $\vert \pm (S-1) \rangle$, and we expect that
roughly $kT_c \sim \Delta \omega /2 \pi$ (in fact $\Delta \omega \sim 11.6K$
($5.1K$) for Mn-12 (Fe-8) respectively, whereas experimentally $kT_c \sim
2K$ ($0.4K$). As noted above, if $T \ll T_c$, everything but the nuclear
subsystem is frozen (dipolar flip-flop processes to levels
$\vert \pm (S-1) \rangle$ occur with frequency
$\sim \Omega_{dip} e^{- \Delta \omega / kT}$, where the coupling
$\Omega_{dip} \sim 10^6-10^8 Hz$ in these systems; at the lowest temperatures
($70 mK$) used in the Fe-8 experiments \cite{Sangregorio}, the activation
exponent was $\sim e^{- 70}$!!). In this case we expect
square root relaxation; and $\tau_Q$ will only depend on temperature to the
extent that $T_2$ does (ie., very little). However as one approaches $T_c$,
molecular flip-flops are more frequent. The square
root prediction is not affected- but
$\tau_Q$ can develop a thermally-activated component, because
(a) now dipolar field fluctuations can also bring molecules to resonance, and
(b) by exciting nuclear $T_1$ transitions, they sweep the nuclear bias (this
mechanism will only be important for strong hyperfine interactions). A detailed
theory of this is quite complicated \cite{JMMM}.

Finally, in the very low-$T$ limit 2 things will happen; the nuclear spins
will polarise in the hyperfine field, for temperatures below the hyperfine
coupling $\omega_)$, and the nuclear $T_2$ fluctuations
will freeze out; and glass theory then predicts that a "dipolar gap" will
open up in $N_{\alpha}(\xi)$ if $M \ll M_0$.

\section{Experiments}

Some of the predictions (such as the square root law
in the quantum regime) have recently been found, in
experiments on Fe-8 and Mn-12 single crystals
\cite{Ohm98,Thomas98}.
However there is still no {\it direct} proof for the
controlling role of nuclear
spins in the quantum dynamics of nanomagnets; the observed
$\sqrt{t/\tau_Q}$ relaxation clearly shows the role of dipole interactions,
but the evidence for the nuclear role is indirect. How can it be seen
directly?

One obvious way
is to do AC absorption experiments over a frequency range
encompassing both $\Delta$ and $\xi_o$. In Mn-12 and Fe-8,
$\Delta$ is perhaps $10^{-10}-10^{-8}K$, some 8-10 orders of magnitude
smaller than $E_D$ or $E_0$
(which is why it is so astonishing that one sees very low $T$ relaxation
at all); we expect $\xi_o \sim 10^{-5}K$. This test is
discussed by Rose and Stamp \cite{Rose}.
A much more intriguing possibility would be available if NMR experimentalists
could find resonance lines for some of the spin-5/2 Mn nuclei in Mn-12
(the zero-field hyperfine coupling is probably $\sim$ 600 MHz (450 MHz) on
the spin-2 (3/2) Mn electronic sites respectively). At this point a "David
and Goliath"
experiment becomes possible; one "tickles" the nuclear spins at the
resonance frequency, in order to control the relaxation of the giant magnetic
molecules (eg. retarding
the relaxation by changing $\xi_o$). Another, probably more difficult
experiment would "freeze out" the nuclear dynamics by cooling to $kT \ll
\omega_o$ (ie, well below $30 mK$ for
Mn-12). The problem here is that of cooling the nuclei, since most of the
heat transfer will involve the same $T_2$ fluctuations that we want to
freeze out! However there is a fundamental interest to doing this, since
according to the theory these nuclear fluctuations are the last barrier to
truly macroscopic coherent (MQC) behaviour in low-$T$ systems (note that
recent work \cite{RPP}
indicates that nuclear and
paramagnetic spins will make it very hard to see
MQC
even in {\it superconductors}; spin baths tend to be far more damaging to
MQC than oscillator baths \cite{Kondo}). This, as well as the resulting
"Quantum Spin Glass" behaviour in the unpolarised limit, means that the
low-$T$ limit promises much to the intrepid experimentalist!

\section{Acknowledgements}

We thank the Russian foundation for basic research (97-02-16548),
INTAS-RFBR 2124, and the CIAR; and P. Nozieres for his hospitality.


\end{document}